\documentclass[jcp,twocolumn,showpacs,preprintnumbers,amsmath,amssymb]{revtex4}
\usepackage{graphicx}
\usepackage{dcolumn}
\usepackage{bm}
\usepackage{epstopdf}
\newcommand{\fig}[1]{Fig.~\ref{#1}}
\newcommand{\be}[1]{\begin{equation}\label{#1}}
\newcommand{\ee}{\end{equation}}

\begin{document}

\title{Atomic and Molecular Systems Driven by Intense Chaotic Light}
\author{Kamal P. Singh$^1$ and Jan M.~Rost $^2$}

\affiliation{$^1$ Department of Physics, Indian Institute of
Science Education and Research Mohali, Chandigarh 160019, India.}
\affiliation{$^2$ Max Planck Institute for the Physics of Complex
Systems, N\"othnitzer Strasse 38, 01187 Dresden, Germany. }

\begin{abstract}
We investigate dynamics of atomic and molecular systems exposed to
intense, shaped chaotic fields and a weak femtosecond laser pulse
theoretically. As a prototype example, the photoionization of a
hydrogen atom is considered in detail. The net photoionization
undergoes an optimal enhancement when a broadband chaotic field is
added to the weak laser pulse. The enhanced ionization is analyzed
using time-resolved wavepacket evolution and the population
dynamics of the atomic levels. We elucidate the enhancement
produced by spectrally-shaped chaotic fields of two different
classes, one with a tunable bandwidth and another with a narrow
bandwidth centered at the first atomic transition. Motivated by
the large bandwidth provided in the high harmonic generation, we
also demonstrate the enhancement effect exploiting chaotic fields
synthesized from discrete, phase randomized, odd-order and
all-order high harmonics of the driving pulse. These findings are
generic and can have applications to other atomic and simple
molecular systems.
\end{abstract}

\pacs{       
}
\maketitle

\section{INTRODUCTION}

The dynamics of atomic and molecular systems exposed to shaped
ultrashort, intense light pulses is a subject of great current
interest \cite{Zewail,Assion,RabitPhysTod}. The research in this
area is motivated by the fact that such light-matter interactions
are mostly nonlinear and nonperturbative in nature, thus offering
an efficient tool to manipulate atomic and molecular phenomenon on
a femtosecond/attosecond time-scale \cite{AttoScience}. A basic
objective of the quantum control is to guide a quantum system from
its initial state towards a desired final state selectively with
high probability \cite{RabitPhysTod}. Several quantum control
strategies exist in the literature such as, using pump-probe
techniques with infrared (IR) and XUV pulses
\cite{Johnsson,AttoSpect}, implementing closed-loop (iterative)
optimization schemes \cite{RabitPhysTod}, and using open-loop
schemes employing tailored light pulses \cite{Rabitz-1}.

There is also some work investigating random light fields as a
mean to control quantum processes such as bond dissociation
\cite{kenny}, photoionization of neutral atoms \cite{kamal_prl} or
Rydberg atoms \cite{NoiseIonRyd}, population transfer between
energy levels of a system \cite{Rabitz-2} and noise
autocorrelation spectroscopy \cite{milner}. This is partly
motivated by the fact that under realistic conditions of the
laser-matter interaction seemingly random fluctuations are
unavoidable. This raises a question about their role on
influencing various quantum processes of interest such as
photoionization, bond-dissociation, and population transfer
between energy levels in a multilevel system etc. Furthermore, it
is known that in certain quantum (as well as classical) systems,
the presence of noise plays a constructive role via the phenomenon
of stochastic resonance (SR) \cite{Gamat,Buchl,QSR}, which can
optimize the system response to a weak driving field. SR requires
three ingredients: a nonlinear system, a coherent driving, and a
noise source \cite{Gamat}. In principle, all three ingredients are
present in atomic and molecular systems interacting with shaped
light pulses. Even very intense quasi-random electric-fields can
be generated with modern pulse shaping techniques \cite{chaoLgt,
MIRshaping, Weiner} which opens up the possibility to employ such
chaotic pulses for atomic and molecular systems.  Given the vast
multitude of pulse shaping possibilities on the one hand side and
the fact that, on the other hand, for a large combined effect of
coherent driving and noise both  must have comparable strength
\cite{kamal_prl}, one needs to identify generic conditions under
which chaotic light (with or without a laser pulse) can serve as a
control tool and assess how efficient various scenarios are.

In the following, we provide a detailed study of the influence of
chaotic light on a single-electron atom interacting with an
ultrashort laser pulse. We will demonstrate the role played by
chaotic light in optimizing the photoionization process and
analyze its properties in the time-domain. After identifying
crucial gain-providing frequency-bands under the driving laser
pulse, we discuss the efficiency of chaotic light generated using
shaped spectral content. In particular, we consider a case when
the chaotic light is generated by discrete frequency components
analogous to the ones produced in the high harmonic generation
process \cite{HHG}. The features discussed here are generic and
can be applied to the problem of molecular dissociation processes.

The article is organized as follows. Section II introduces our
model of an atomic system interacting with a femtosecond IR laser
pulse and a chaotic light pulse. The method to solve the
corresponding time-dependent Schrodinger equation (TDSE) along
with useful physical observable is explained. In section III, we
show how the chaotic light can produce optimal photoionization
enhancement. To characterize the enhancement mechanism, we compute
time-resolved wavepacket dynamics and a frequency-resolved gain
profile of the driven-atom. We then consider spectrally shaped
chaotic lights to optimize the photoionization, in particular, the
ones generated by phase randomized even-order and all-order higher
harmonic components. Finally, section IV provides a summary of
results with our conclusions.

\section{DESCRIPTION OF THE MODEL}

\subsection{Atomic and Molecular system exposed to chaotic light and a laser pulse}
Let us consider a generic example of an atomic (or molecular)
system with one degree of freedom such as the simplest
single-electron hydrogen atom. Due to the application of an
intense linearly polarized laser field $F(t)$, the electron
dynamics is effectively confined in one-dimension along the laser
polarization axis \cite{Eberly}. The Hamiltonian for such a
simplified description of the hydrogen atom is written as (atomic
units, $\hbar=m=e=1$, are used unless stated otherwise),
 \begin{equation}  
   H(x,t) = H_0(x) + x F(t) + x Z(t),
   \label{eqn:eqn1}
 \end{equation}
where $x$ is a system coordinate such as the position of the
electron and $H_0(x)$ is the unperturbed Hamiltonian. The external
perturbations, a laser pulse $F(t)$ and a chaotic field $Z(t)$,
are dipole-coupled to the system. In the case of an atomic system,
the unperturbed Hamiltonian is given by $H_0(x)=
\frac{\hat{p}^2}{2} + V(x)$. The potential $V(x)$ is approximated
by a non-singular Coulomb-like form \cite{Eberly},
 \begin{equation}  
     V(x)=- \frac{1}{\sqrt{x^2 + a^2}}.
   \label{eqn:eqn2}
 \end{equation}
Such a soft-core potential with parameter $a$ has been routinely
employed to study atomic phenomenon in strong laser fields
\cite{Eberly, mpi}.

Note that a similar one-dimensional model can be used to study the
photo-dissociation of diatomic molecules. The potential in that
case is the Morse potential. Such model has been used to study for
example the molecular dissociation by white shot noise
\cite{kenny}, and using a combination of a white noise with a
mid-IR laser pulse \cite{kamal}. The results on the role of
chaotic light on the atomic system can be translated, at least
qualitatively, to the case of dissociation dynamics of the
diatomic molecules.

 \begin{figure}[t]   
    \includegraphics[width=.9\columnwidth]{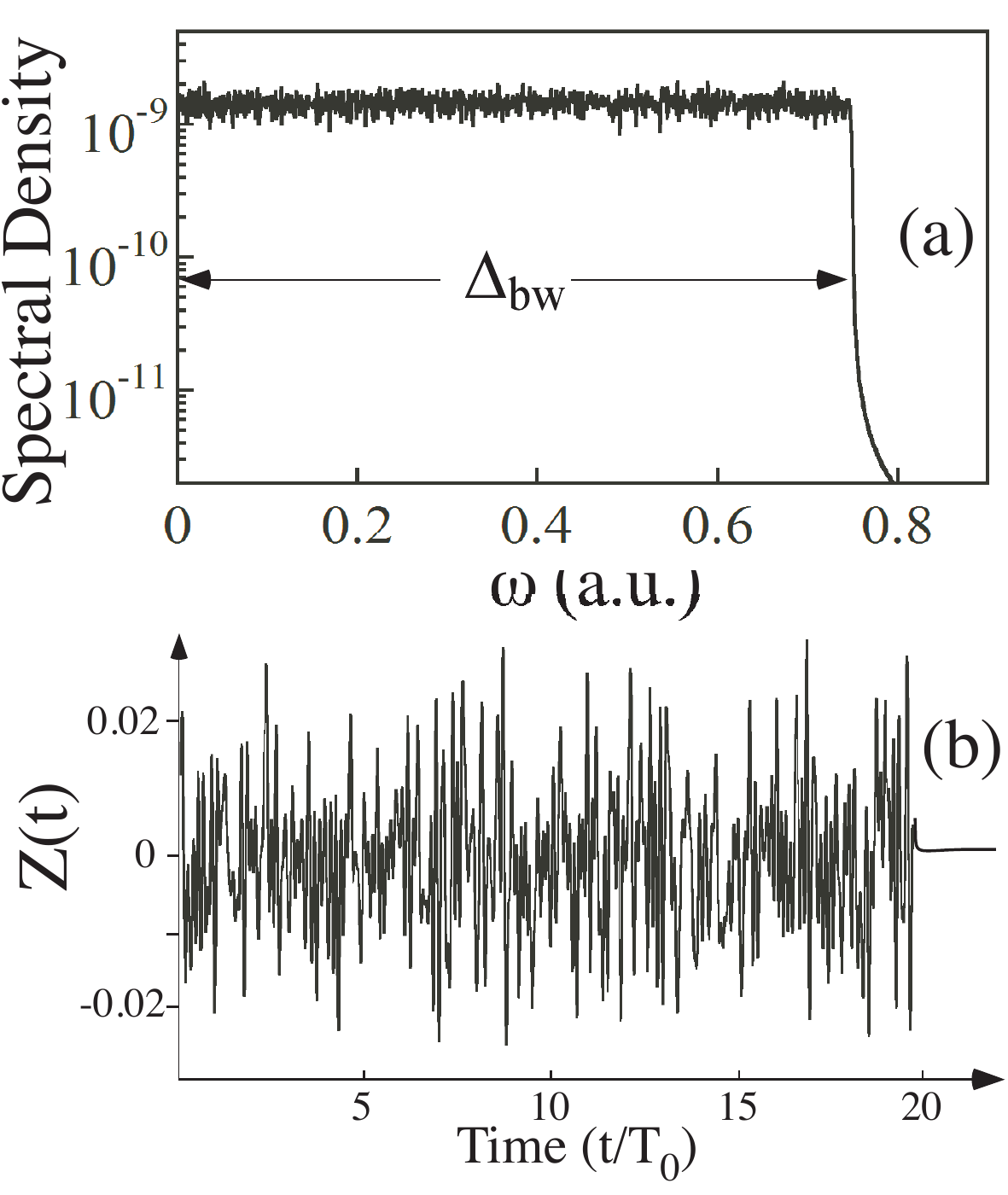}
 \caption{
The principle of chaotic light generation. (a) The power spectral
density (PSD) of the chaotic light defining its bandwidth
$\Delta_\mathrm{bw}$. (b) A typical realization of the chaotic electric
field $Z(t)$ of bandwidth $0.75$, $N=1024$.
  }
\end{figure}

The laser field in Eq.~1 is a nonresonant mid-infrared (MIR)
femtosecond pulse described as,
 \begin{equation}  
 \mbox{$ F(t) = f(t) F_0\sin(\omega_0 t+\delta)$}.
 \label{Lsrpulse}
 \end{equation}
Here $F_0$ defines the peak amplitude of the pulse, $\omega_0$
denotes the angular frequency, and $\delta$ is the
carrier-to-envelop phase. We choose a smooth pulse envelop
$f(t)=\sin^{2}(\pi t/T_p))$ where $T_p$ is the pulse duration.

The chaotic field term $Z(t)$ consists a normalized sum of $N$
closely spaced oscillators defined as \cite{kamal_pra},
\begin{equation}            
 Z(t) = \sqrt{\frac{2}{N}} \sum_{n=1}^N g(t) F_\mathrm{rms}\sin(\omega_n t + \phi_n),
 \label{eqn:eqn4}
 \end{equation}
where $\omega_n$, $\phi_n$ denotes the angular frequency, phase of
$n^{th}$ mode, and $F_\mathrm{rms}$ is the root-mean-square amplitude of
$Z(t)$. The chaotic light envelop $g(t)$ is chosen such that
$g(t)=1$ for $0 \leq t \leq T_p$ and zero elsewhere. The spacing
between two modes $\delta \omega = \omega_{n+1} -\omega_n$ is kept
very small, so as to have a quasi continuous frequency spectrum
[see Fig.~1(a)]. One can define a characteristic bandwidth (BW) of
the chaotic light as, $\Delta_\mathrm{bw}=N \delta \omega$. Note that
here we consider these frequency modes to oscillate independently
of each other with their phases $\phi_n$ assuming random values
relative to each other. In this particular case of
phase-randomized coherent modes, the averaged field $Z(t)$ at any
point will be noise-like as shown in Fig.~1(b), hence the name
chaotic field. Various realizations of the chaotic light are
generated by generating different sets of $\phi_n$ with uniform
probability distribution between 0 to $2\pi$. Note that $Z(t)$
becomes a white noise term in the limit $\Delta_\mathrm{bw} \to \infty$
and $\delta \omega \to 0$.

\subsection{Quantum dynamics and physical observable}
We solve the time-dependent stochastic Schr\"odinger equation,
 \begin{equation}    
   i\; \frac{\partial \Psi(x,t)}{\partial t} = H(x,t)\; \Psi(x,t),
 \label{eqn:tdse}
 \end{equation}
driven by a chaotic light field and a laser pulse. Due to the
presence of the chaotic field term in the Hamiltonian as described
above, the quantum evolution is different for each realization of
$Z(t)$. This requires to average the final observable of interest
over a large number of realizations of $Z(t)$. For a given
realization, the numerical solution of the Schr\"odinger equation
amounts to propagating the initial wave function $\vert \Psi_0
\rangle$ by computing the infinitesimal short-time stochastic
propagator, using a standard split-operator fast Fourier algorithm
\cite{SOper}.

Note that the initial state $\vert \Psi_0 \rangle$ is always
chosen to be the \emph{ground state} of the system 
with an energy of $I_b = -0.5$~a.u.. This is obtained by the
imaginary-time relaxation method for $a^2=2$ \cite{Eberly}.  To
avoid parasitic reflections of the wavefunction from the grid
boundary, we employ an absorbing boundary \cite{SOper}.

The main observable of interest for us is the total ionization
probability defined as,
 \begin{equation}  
  P = 2\int_{0}^{\infty} J_R(x_R,t) dt.
 \label{eqn:IP}
 \end{equation}
Here $J_R(x_R,t) = Re\lbrack \Psi^\ast \: \hat{p} \: \Psi
\rbrack_{x_R}$ is the ionization flux \cite{QMbook} leaking in the
continuum where $x_R$ is a point near the absorbing boundary. In
the following sections, we shall show how the combination of a
chaotic field and the laser pulse can enhance photoionization.

\section{RESULTS AND DISCUSSIONS}
\subsection{Optimal enhancement of photoionization by chaotic fields}

\subsubsection{A laser pulse induced femtosecond photoionization}
\begin{figure}[t]
  \begin{center}
    \includegraphics[width=.9
    \columnwidth]{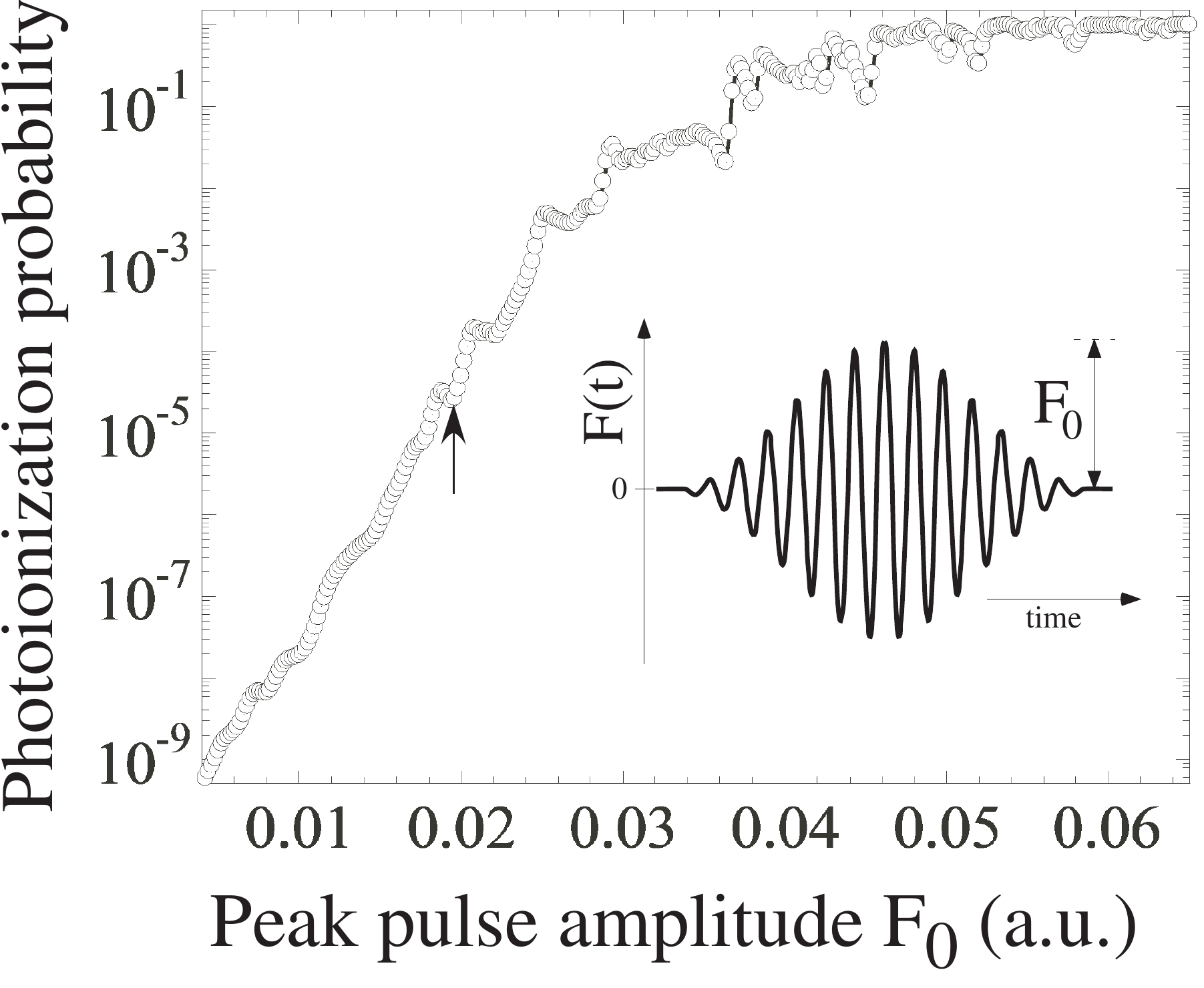}
    \caption{
The ionization probability versus the laser peak amplitude $F_0$.
Inset: schematic of the laser pulse. A vertical arrow on the curve
corresponds to a weak laser pulse amplitude, $F_0=0.02$. }
\label{fig:fig2}
   \end{center}
\end{figure}
We shall first consider the atomic photoionization due to a short
but strong laser pulse only. As one varies the peak amplitude
$F_0$ of the mid-IR laser pulse ($ \omega=0.057$), the ionization
probability first increases nonlinearly, and then saturates to the
maximum value of unity, for $F_0>0.05$. This behavior of
$P_l(F_0)$ is a characteristic signature common to many atomic (or
molecular) systems exposed to intense laser pulses \cite{mpi}.

The laser pulse produces (nonlinear) ionization of the atom which
is most easily understood with the picture of tunneling
ionization. Ionization flux is produced close to times $t_{k}= [(k+\frac 12)\pi-\delta]/\omega_{0}$ when
the effective potential \mbox{$U(x,t)=V(x)+x F(t)$}, is maximally
bent down  by the dipole-coupled laser field. One can therefore
conclude that the photoionization dynamics is highly nonlinear,
and in particular it exhibits a form of ``threshold'' which is defined
 by the condition for over barrier ionization,
$I_{b}=\mathrm{max}_{x}U(x,t_{k})$.
From Fig.~2 we identify a ``weak'' amplitude laser pulse around
$F_0=0.02$ (see arrows in Fig.~2) such that it produces almost
negligible ionization probability. In the following, we shall
study the effect of adding a chaotic field on the photoionization
process.

\subsubsection{Enhancement by addition of the chaotic light}

\begin{figure}[t]    
\includegraphics[width=.9\columnwidth]{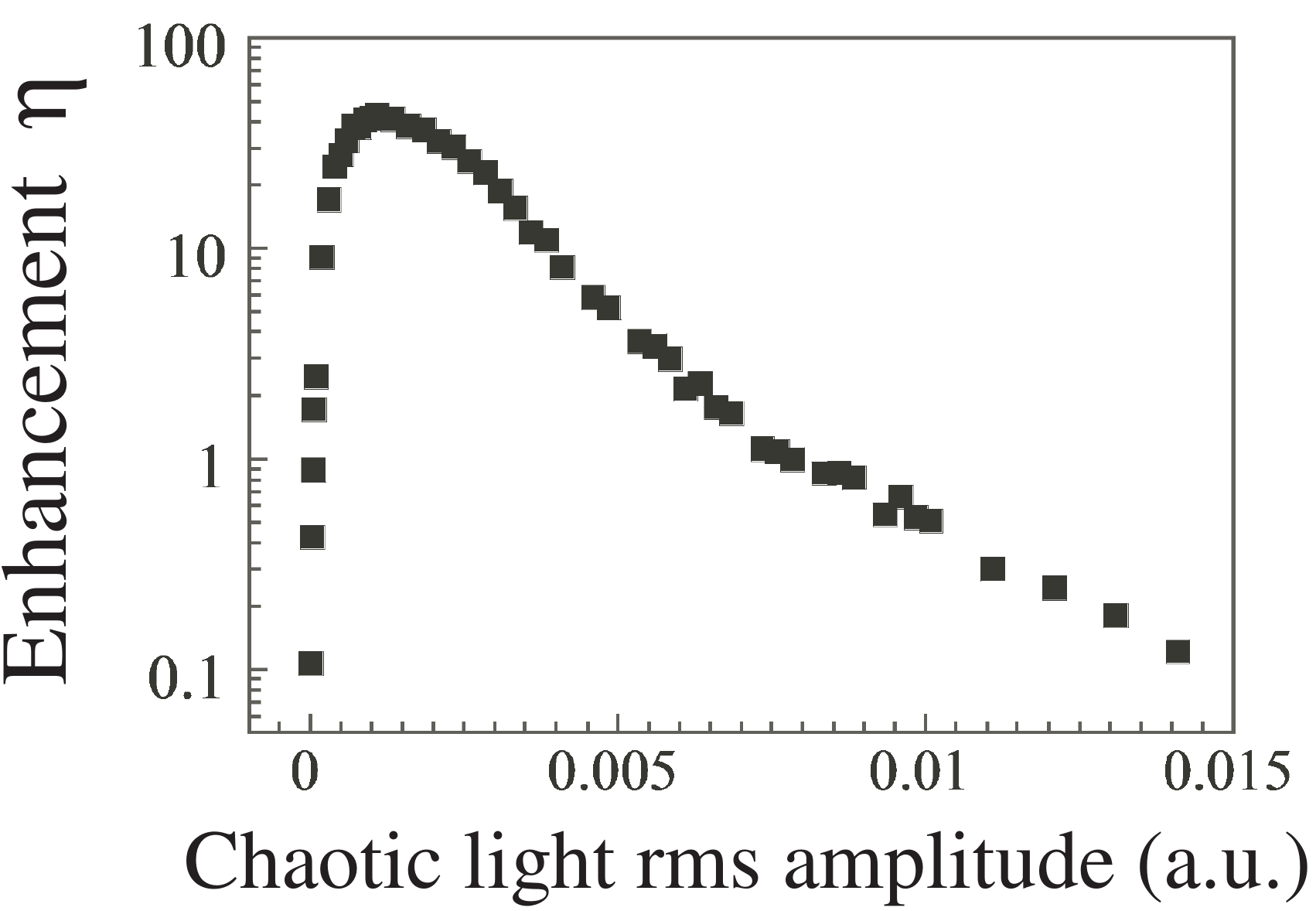}
\caption{The factor of enhancement $\eta$ plotted versus the rms
amplitude of the chaotic light ($\Delta_\mathrm{bw}=0.75$, $N=512$). At
the maximum of the curve the ratio $F_\mathrm{rms}/F_0 = 0.075$. The
curve is averaged over 50 different realizations of the chaotic
light.
 }
\label{fig:fig3}
\end{figure}

Here we look into the possibility of efficiently ionizing the
atom, when it is subjected to both a weak laser pulse ($F_0=0.02$)
and chaotic light. In order to characterize the stochastic
enhancement in the atomic ionization due to interplay between the
chaotic field and the laser pulse, we use the following definition
of the enhancement factor  as \cite{kamal}
 \begin{equation}            
 \eta = \frac{P_{l+n} - P_0}{P_0}
 \label{eqn:eqn6}
 \end{equation}
where $P_l$ is the ionization probability due to laser pulse
alone, $P_n$ is IP due to chaotic light alone, and $P_{l+n}$ is
the one due to the combined action of both the pulses. One can
verify that a zero value of $\eta$ corresponds to the case when
either the laser pulse ($P_l \gg P_n$) or the noise ($P_l \ll
P_n$) dominates. Furthermore, $\eta$ characterizes a truly
nonlinear quantum interplay as it also vanishes if we assume a
``linear'' response as a sum of individual probabilities, $P_{l+n}
= P_l + P_n $.

In Fig.~3, we have plotted the enhancement factor $\eta$ versus
the rms amplitude $F_\mathrm{rms}$ of a broadband chaotic field
($\Delta_\mathrm{bw}=0.75$). Each data point is averaged over 50
different realizations of the chaotic light. As the $F_\mathrm{rms}$
increases, $\eta$ exhibits a sharp rise, followed by a maximum at
a certain value of the (optimum) noise, and then a gradual fall
off for very intense chaotic fields. It is worth mentioning that
only a modest noise-to-laser ratio ($F_\mathrm{rms}/F_0 = 0.075$) is
required to reach the optimum enhancement (here $\eta_\mathrm{max} =
36$). Note that such a curve has been obtained by white Gaussian
noise \cite{kamal_prl}. We have verified that the location of the
optimum enhancement is governed by an empirical condition, where
the strengths of the laser pulse and noise are comparable such
that $P_l \sim P_n $.

The enhancement curve versus the chaotic light amplitude bears
striking resemblance to the typical SR curve where the system
response is plotted versus the noise amplitude. However, the
observed effect is not the standard SR effect with the
characteristic time-scale matching between coherent and incoherent
driving. The underlying gain mechanism here is completely
different, as we shall see later. In the sense of the existence of
an optimum noise level, one can perhaps call this as a generalized
quantum SR for such atomic systems on ultrafast time-scales.

\begin{figure}[t]    
\includegraphics[width=.9\columnwidth]{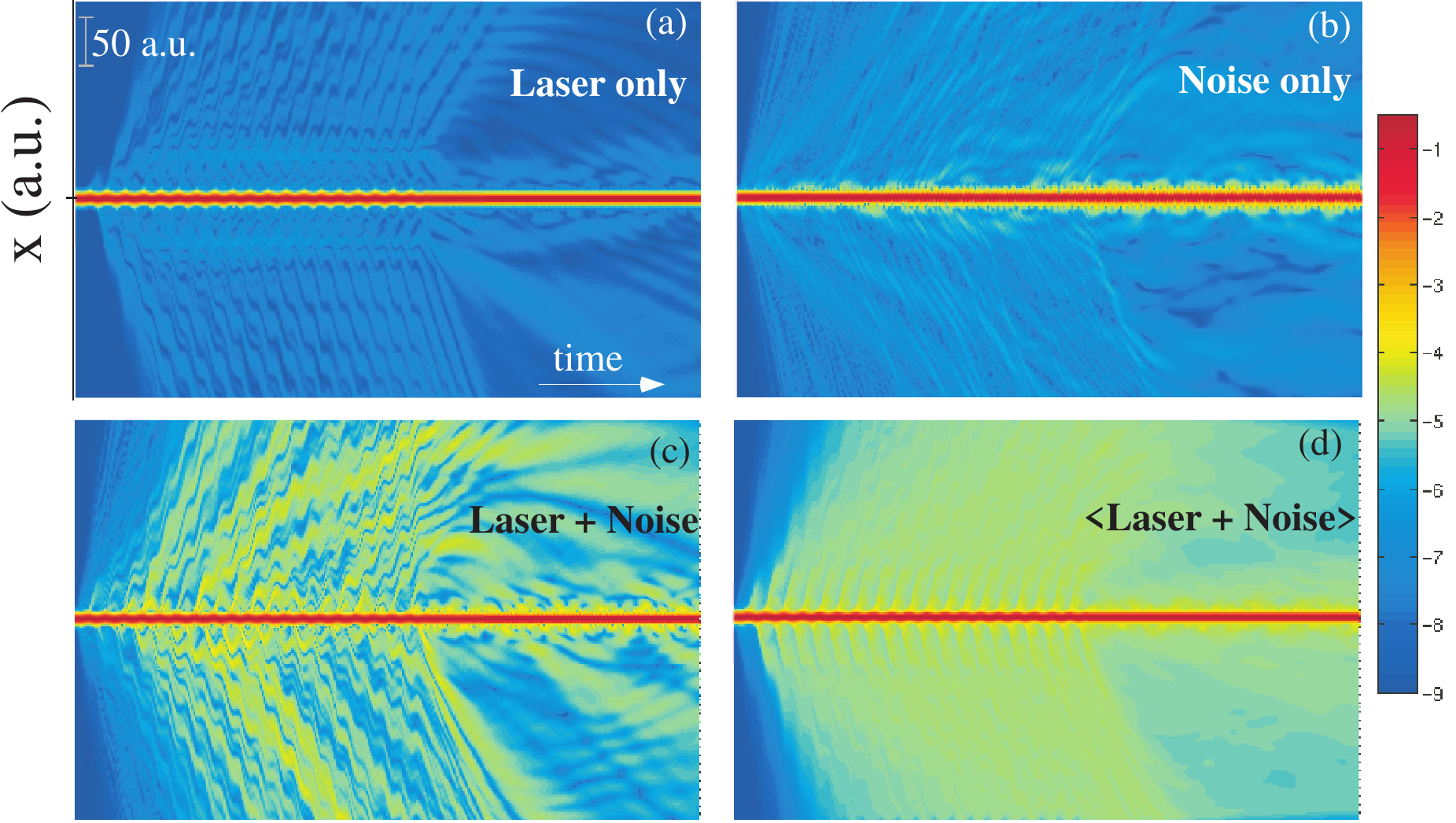}
\caption{Temporal evolutions of the probability density at the
optimal enhancement point of Fig.~3. (a) The case of laser pulse
only ($F_0=0.02$), (b) Chaotic light only ($F_\mathrm{rms}=0.016$), (c) a
combination the chaotic light and the laser pulse, (d) same as in
(c) but ensemble averaged over 50 different realizations of the
chaotic light.
 }
\label{fig:fig4}
\end{figure}

\subsection{Analysis of the optimal photoionization enhancement}

\subsubsection{Time-resolved wavepacket dynamics and level populations}
We will now analyze the mechanism of the stochastic enhancement in
the time domain. To this end, we compute the wavepacket dynamics
and the level populations at the optimum of the enhancement curve.
As shown in Fig.~4(a), the weak laser pulse does not yield any
significant ionization. Similar is the case when the atom is
subjected to the chaotic light alone p[Fig.~4(b)]. The chaotic
light induced wavepacket leaking happens at arbitrary instances in
time when compared to the one for laser pulse where it takes place
regularly near the pulse maxima and minima. When we combine the
laser pulse and chaotic light, the ionization flux becomes
significantly enhanced, as is clearly visible in Fig.~4(c). Note
that the color scale is logarithmic here. Due to the combined
action of the laser pulse and noise, significantly higher fraction
of the wavepackets is leaked to the continuum. These wavepackets
drift away to the continuum and contributes to the enhanced
ionization probability. Fig.~4(d) shows the probability density
evolution averaged over 50 different realizations of the chaotic
light. This ensemble averaging washes out the finer details of the
wavepacket dynamics, however, it shows a more homogeneous
enhancement of the ionization flux. This shows that one can
clearly visualize the photoionization enhancement with wavepacket
dynamics.

 \begin{figure}[t]   
    \includegraphics[width=.9\columnwidth]{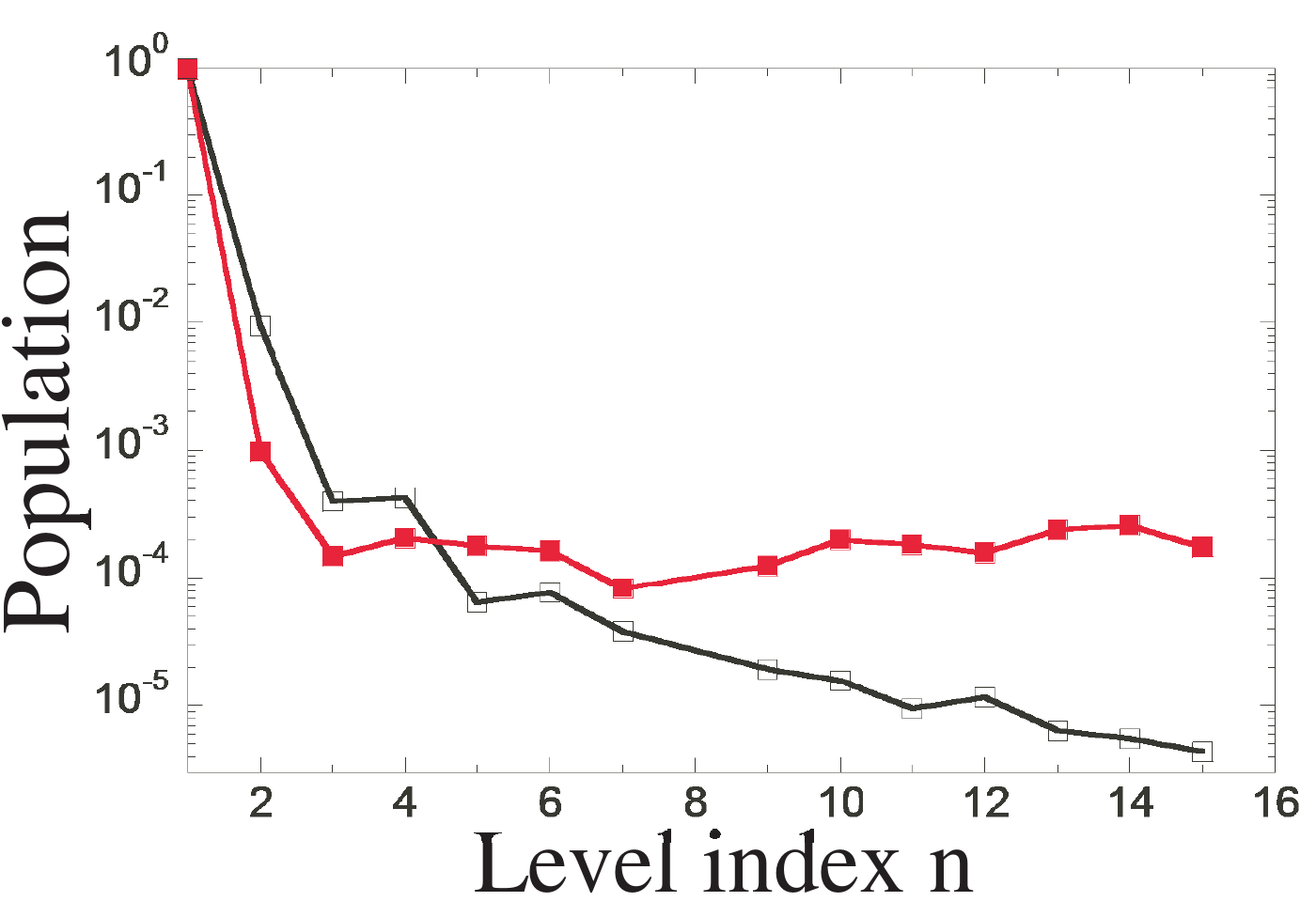}
 \caption{
The populations of first fifteen energy levels after the atom has
interacted with light pulses corresponding to the optimum of
Fig.~3. Filled square: chaotic pulse only $F_\mathrm{rms}=0.016$, and
open square: The chaotic light with the weak laser pulse,
$F_0=0.02$.
  }
\end{figure}

How does this enhancement manifest itself in the level
populations? To answer this, we compute the populations of the
first $15$ unperturbed atomic energy levels at the end of the
laser pulses. In the case of chaotic light only, the populations
of the excited states is small and it decreases as the level index
increases (see Fig.~5). However, in the case of simultaneous
actions of the laser and chaotic light, the higher excites states
populations are significantly enhanced, as one can see in Fig.~5.
The population dynamics as well as the wavepacket dynamics clearly
show the enhanced excitation (ionization) as a result of the
simultaneous action of the laser pulse and noise.

\subsubsection{Frequency-resolved gain profile}
In order to understand better how the chaotic light enhances
ionization in combination with an IR driving, we need to identify
the frequency components through which the driven system can
absorb energy from the chaotic light since these are not just the
atomic lines due to the strong driving.  To find out where the
driven system absorbs efficiently, we compute a frequency-resolved
atomic gain (FRAG) profile $G(\nu)$ using a pump-probe type of
setting as described below.
 \begin{figure}[t]    
 \includegraphics[width=.9\columnwidth]{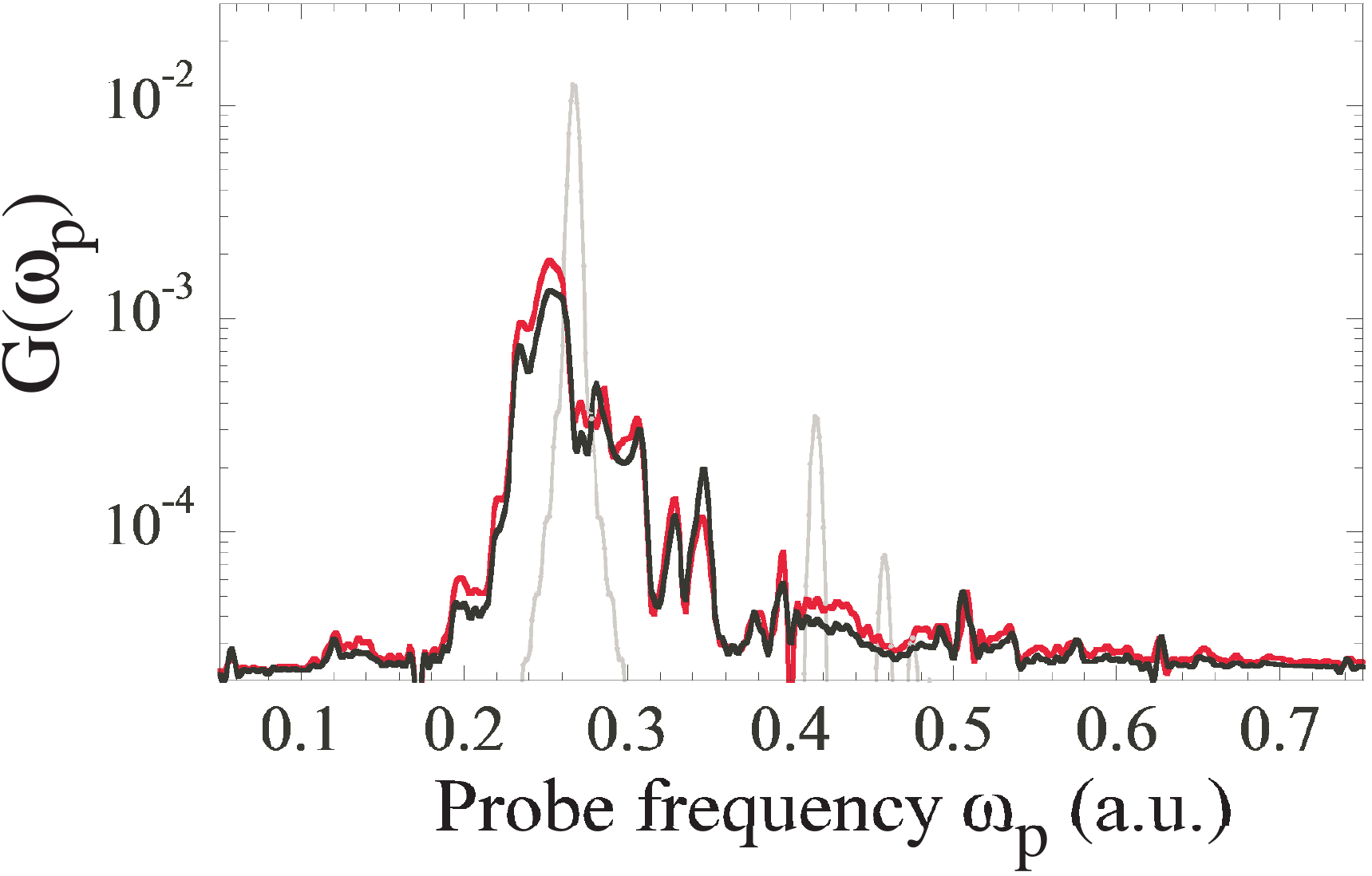}
 \caption{ Frequency-resolved gain profile of the atom which is
driven by a 10 cycle laser pulse $\omega = 0.057$ of amplitude
$F_0=0.02$. The atomic gain is probed by a synchronized
application of a weak probe beam $F_p = 0.0005$ of tunable
frequency $\omega_p$. Both the depletion of the ground state
population (thick red line) and the net absorption of energy
(black) due to the probe are shown. The gain of the bare atom
(thin gray line) is also shown for comparison.
 }
 \label{fig:fig6}
 \end{figure}

We simply replace the chaotic light by a tunable monochromatic
probe pulse, \mbox{$ F_p(t) = f(t) F_p \sin(\omega_p t)$}. The
amplitude $F_p$ of the probe is much weaker than the driving laser
pulse [$F_p/F_0 = 0.025$], such that it does not significantly
alter the quasi energy levels, but can drive the transition
between them. By preparing the atom in its ground state and
measuring the depletion of its population (which indicates the
absorbed energy) as a function of the probe frequency $\omega_p$,
a FRAG profile of the atom is obtained.

Such a gain profile $G(\nu)$ is shown in Fig.~6 for the 10 cycle
long laser pulse of amplitude $F_0=0.02$. Compared to the pure
atomic transitions (thin line) the FRAG shows a complicated
structure for the driven atom. The dominated absorption is the
slightly red shifted atomic transition frequency between the
ground and first excited state, $\omega^\mathrm{shifted}_{12}$,
but strongly broadened. In fact the broadening is so strong that
the driven atom absorbs in the entire frequency regime from
$\omega^\mathrm{shifted}_{12} \approx 0.25$) to the full binding
energy ($\omega_{c} = 0.5$). This will have consequences for the
enhancement of ionization  of the driven atom under chaotic light
as we will see below. In particular, we shall employ different
variants of chaotic light generated by a variety of spectral
shapes to study these consequences in detail.
 \begin{figure}[t]    
 \includegraphics[width=.9\columnwidth]{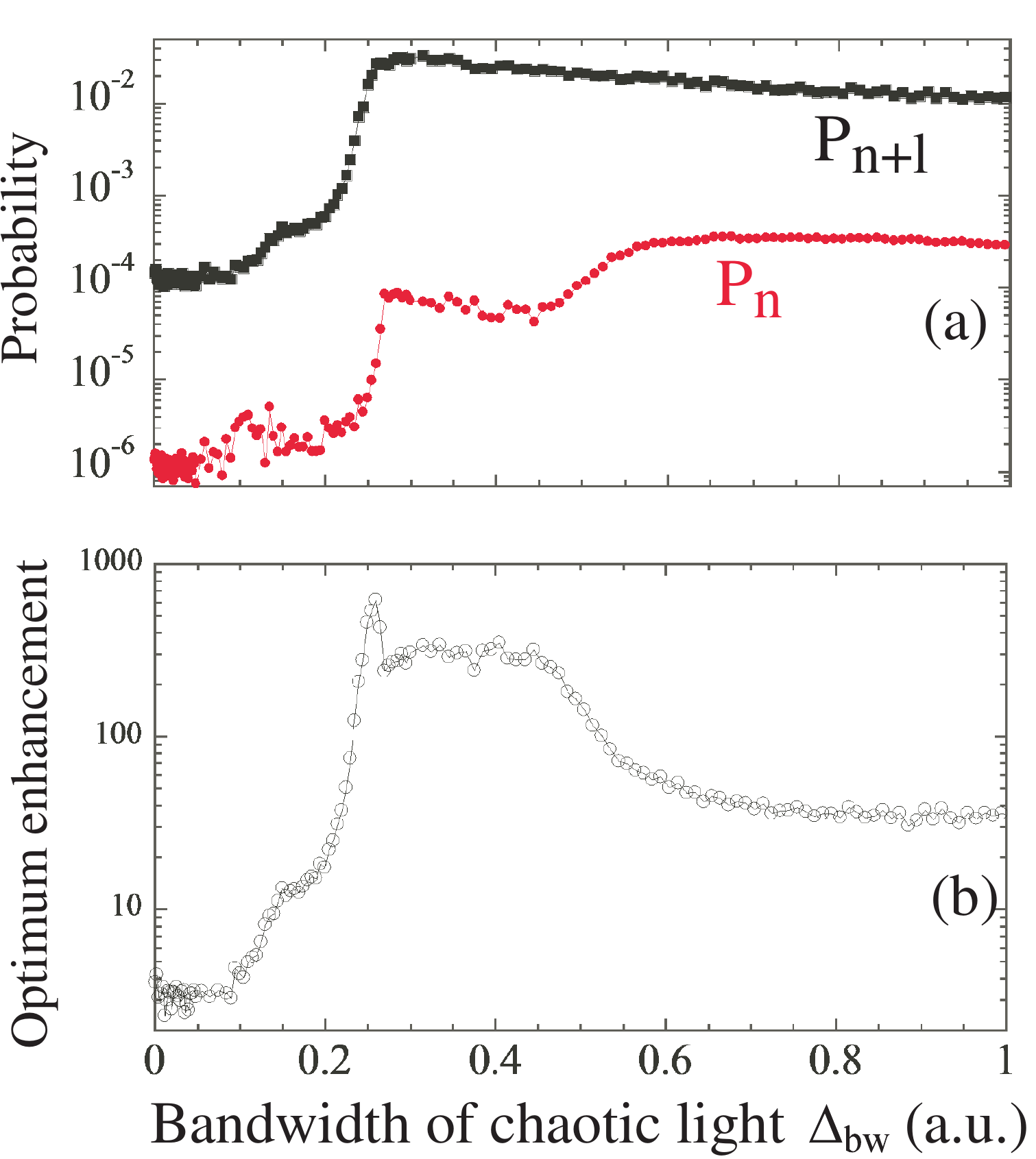}
 \caption{
The role of tuning the chaotic light bandwidth $\Delta_\mathrm{bw}$ on
the optimum enhancement ($F_0=0.02$, $F_\mathrm{rms}=0.016$). (a)
Ionization probabilities versus $\Delta_\mathrm{bw}$ for chaotic
light only, $P_n$, and for a combination of chaotic light and
laser pulse, $P_{n+l}$. (b) A corresponding enhancement profile
$\eta$ obtained from these probabilities versus
$\Delta_\mathrm{bw}$.
 }
 \label{fig:fig7}
 \end{figure}

 \subsection{Enhancement with spectrally shaped chaotic light }

\subsubsection{Chaotic light of variable bandwidth}
So far we have considered a broadband, quasi-continuous spectrum
of the chaotic light. In the following, we shall elucidate the
role of different types of spectral shapes of the chaotic light on
the ionization enhancement. Firstly, we will vary the  bandwidth
$\Delta_\mathrm{bw}=[0,\omega_\mathrm{max}]$ of the chaotic light
by lowering the higher cut-off frequency $\omega_\mathrm{max}$.
The effect of different $\Delta_\mathrm{bw}$ on the ionization
probability for the chaotic light alone ($P_n$) and for its
combination with the laser pulse ($P_{l+n}$) is shown in Fig.~7.
One can easily understand the curves under the condition that
firstly, only one photon of the noise source can be absorbed, yet
with any frequency within the bandwidth, and secondly, that the
driven system absorbs mainly between the red-shifted transition
frequency $\omega^\mathrm{shifted}_{12}$ and the cut-off frequency
(ionization potential) $\omega_{c}=0.5$ as described above. $P_n$
and $P_{l+n}$ exhibit a sharp onset once the bandwidth of the
chaotic light covers first transition $\omega_{12}\approx 0.267$.
However, a closer inspection reveals that $P_{l+n}$ rises a bit
earlier than $P_{n}$  which is due to the fact that under the
driving the transition frequency $\omega_{12}$ is red-shifted to
$\omega^\mathrm{shifted}_{12}\approx 0.25$ (see \fig{fig:fig6}).
Hence, in the optimum enhancement curve (Fig.~7b), there is an
overshooting resulting in a sharp peak at
$\omega^\mathrm{shifted}_{12}$. From $\omega_{12}$ towards higher
frequencies, $P_{n+l}$ decreases only slightly. This is also true
for $P_{n}$ which, however,  ``shuts off'' with a soft step at the
cut-off frequency of $\omega_{c}=0.5$. This is clear from the fact
that beyond $\omega_{c}$ the driven atom does not absorb energy
(see \fig{fig:fig6}). The slight decrease of both curves also
traces the decreasing availability of the driven atom to absorb
light from $\omega_{12}\approx 0.25$ to $\omega = 0.5$. For the
enhancement this implies overall a plateau between $\omega_{12}$
and $\omega_{c}$ with a sharp onset at $\omega_{12}$ and a soft
offset at $\omega_{c}$ to eventually approach the white noise
limit for infinite bandwidth. This dependence of enhancement on
the bandwidth of the chaotic light suggests that one should use a
broad enough bandwidth [covering the plateau of Fig.~7(b)] in
order to optimize the enhancement.

\subsubsection{Role of quasi-resonant narrow-band chaotic field}
The FRAG curve of Fig.~6 shows that the most interesting region in
the spectral profile is around the $\omega_{12}$ where one sees a
lot of structure. Now we want to consider a special case of system
specific narrow band chaotic light centered at the first resonant
transition, and widen its bandwidth keeping its central frequency
fixed at the $\omega_{12}$. For extremely narrow bandwidth, this
essentially becomes a monochromatic light. This example shall also
compare the enhancements produced using a single frequency
resonant pulse versus the narrowband chaotic light. In Fig.~8 we
plot the enhancement curve as a function of rms amplitude of the
chaotic light. In the case of very narrow BW, which means that the
chaotic light is almost monochromatic, centered at the atomic
resonant frequency, one see the enhancement curve. As we increase
the BW keeping its central frequency unchanged, the enhancement
becomes stronger. This means that although a single frequency
laser can provide the photoionization enhancement, the use of
broadband chaotic light is preferred if one wants to have a large
enhancement factor. With these two cases of narrowband chaotic
light and the BW tunable chaotic light, we not only highlight the
significance of the FRAG but also show how the chaotic light can
be spectrally tuned to a specific system in order to optimize the
enhancement profile. The advantage of using a broadband source is
that the design becomes independent of both the particular atom
and the pulse parameters.
 \begin{figure}[t]    
 \includegraphics[width=.9\columnwidth]{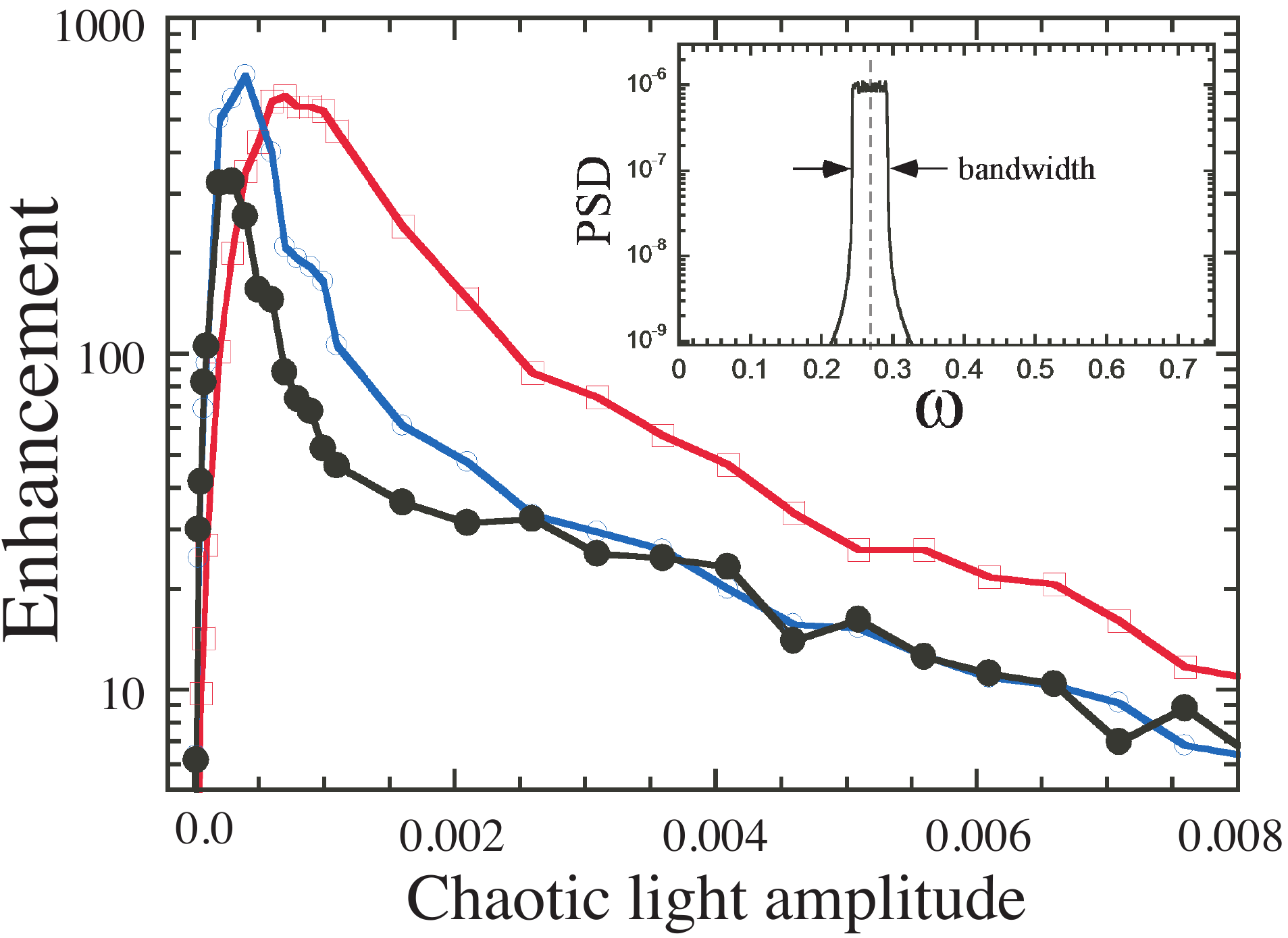}
 \caption{Enhancement induced by a narrow band chaotic light centered at
first atomic transition $\omega_{12}=0.267$ as shown in the inset.
The enhancement curves for the bandwidths, filled circles: 0.001
(black), open circles: 0.015 (blue), and open squares: 0.2 (red).
 }
 \label{fig:fig8}
 \end{figure}

\subsection{Employing chaotic light generated from discrete harmonics of the driving pulse}
It is known that high harmonics (HH) can generate light of
extremely broad bandwidth, albeit in the form of discrete
frequencies which are integer multiples of the driving laser
frequency \cite{HHG}. The relative phases of these harmonics are
either decided in the generation process itself or in the
propagation medium. A great deal of efforts has been put into
locking their phases to generate attosecond pulses
\cite{AttoSpect}. Here we consider the completely opposite case,
where the phases are randomly distributed. This can be realized in
principle via their propagation in a suitably chosen medium. For
phase randomized HH the net electric field varies almost randomly
thus mimicking the chaotic light to some extent. One may wonder if
it is possible to use such a discrete set of frequencies to
observe the enhancement effect. In the following we consider two
different cases where the HH contain, (i) only odd order harmonics
of the driving field and, (ii) both even and odd order components.
Both these cases are experimentally feasible, the only challenge
is to devise an strategy to keep their phases randomized.
 \begin{figure}[t]    
 \includegraphics[width=.9\columnwidth]{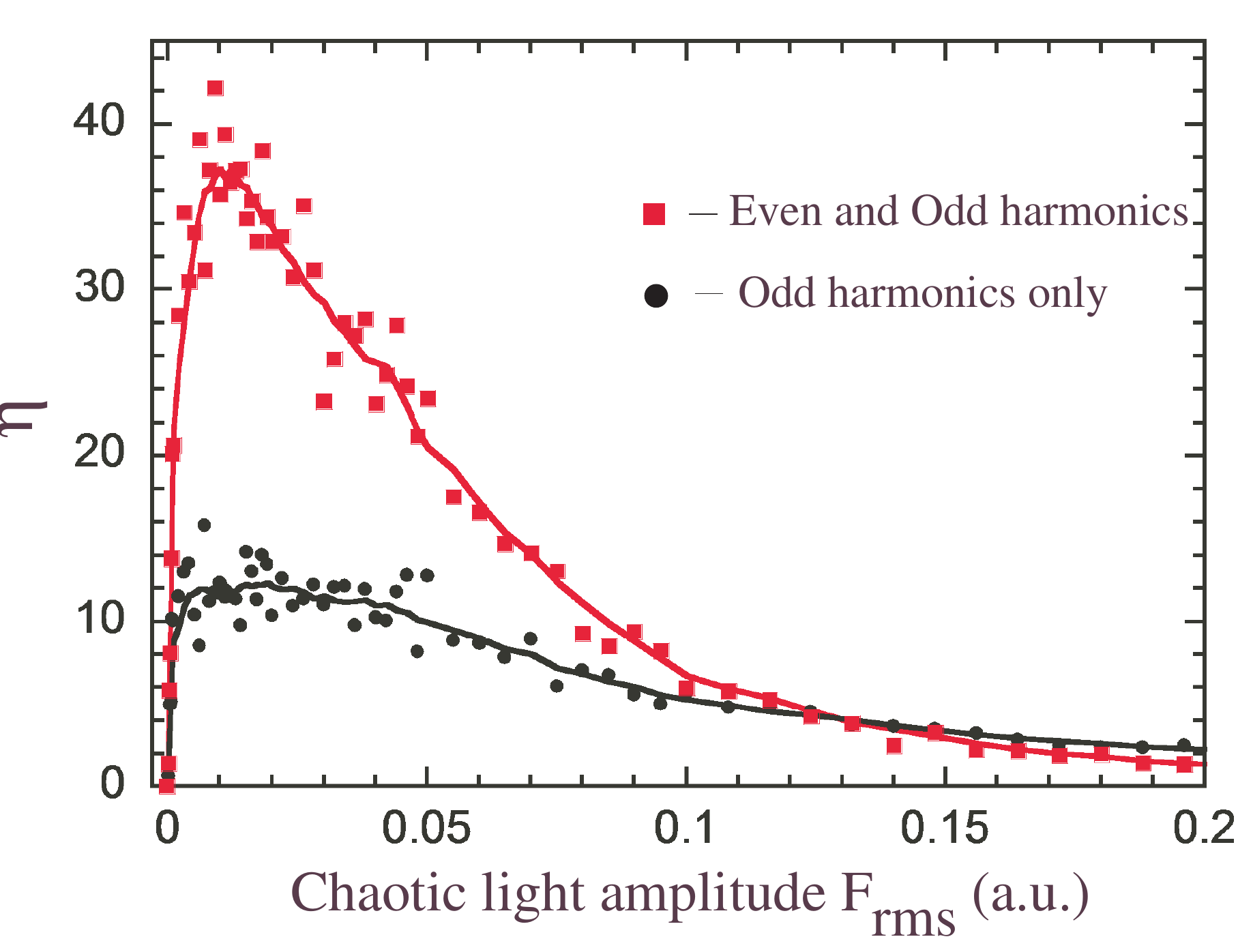}
 \caption{ The enhancement curve for a chaotic light with discrete
frequency spectrum consisting of $n=1,2,\ldots,6$ higher harmonics
of $\omega_{0}=0.057$. Odd-harmonics only: circles, and both even
and odd harmonics: squares. Solid lines are average smoothing
curves. In both the cases six harmonics ($N=6$) are considered
starting from the third one. Each point is averaged over 10
chaotic light realizations.
 }
 \label{fig:fig9}
 \end{figure}
 \subsubsection{Use of only odd-order harmonics}
Considering the case of phase randomized six odd-order harmonics
$\delta \omega = 2\omega_0$, we assign their phases a random value
with a uniform probability density between 0 and $2\pi$. Note that
due to equally spaced harmonics the chaotic light time-series
would repeat itself after half a laser period. The enhancement
curve obtained by averaging over 10 realizations of the HH light
is shown in Fig.~9. This suggests an alternative possibility to
observe enhancement and underscores the generic nature of the
effect, although the maximum enhancement that can be reached in
this way is only around $\eta_\mathrm{max}=14$. The reason for the
relative small enhancement lies again in the FRAG (\fig{fig:fig6})
since the main absorption at $\omega_{12}^\mathrm{shifted}=0.25$
is exactly in the gap of the  odd HH used here, namely between
$3\omega_{0}=0.171$ and $5\omega_{0}=0.285$. One expects higher
enhancement if this gap is filled by all order harmonics.

\subsubsection{Use of all-order harmonics}
We now consider the possibility of creating nonlinear enhancement
of ionization  using chaotic light generated from all-order
harmonics $\delta \omega = \omega_0$. Due to half the frequency
spacing compared to odd harmonics only the intrinsic periodicity
of the chaotic light is now a full optical cycle. Such an HH
spectrum can be experimentally generated by using a two color
driving field, by combining the fundamental field by its second
harmonic \cite{HHG}. As one can see in \fig{fig:fig9}  all-order
HH lead to a stronger enhancement with a maximum of around 35
compared to using odd harmonics only, although the total band
width covered with the 6 all order harmonics is only half that of
odd order harmonics (up to $\omega_\mathrm{max}=8\omega_{0}$ compared to
$\omega_\mathrm{max}=13\omega_{0}$).  What really makes the difference is
that the all-order HH light has with $4\omega_{0}=0.228$ a
frequency component close to the optimum absorption near
$\omega_{12}^\mathrm{shifted}$.

\section{Summary and Conclusion}
We have investigated the role of a combined action of a
femtosecond laser pulse and chaotic light on a generic quantum
phenomenon such as photoionization. Due to the inherent
nonlinearity of the strong field ionization process, a dramatic
enhancement in ionization yield is observed when a modest amount
of optimum chaotic light is added to the weak laser pulse. The
mechanism of the enhancement is analyzed using wavepacket
evolution as well as level population dynamics.

The key to understanding the enhancement is the frequency-resolved
atomic gain profile under the laser pulse which differs substantially
 from the gain profile of an isolated atom. The
ionization enhancement depends on how efficiently the chaotic
light supplies strong gain frequencies of the laser-atom system.
To elucidate  the enhancement we have employed standard
broad band chaotic light and beyond it, various shaped chaotic fields such as
two different classes of variable bandwidth, one with a variable
upper frequency limit and one with a variable bandwidth centered
around the main absorption frequency of the laser-atom system.
Furthermore, motivated by the extremely broad bandwidth offered by
high harmonics, we synthesized chaotic light from phase randomized
discrete odd-order harmonics and all-order harmonics. Such chaotic
fields also lead to an enhancement profile which suggests an
experimental feasibility provided one can randomize the harmonic
phases.

We emphasize that the features observed here are of generic
nature. Analogous effects can be therefore expected in other atomic and
molecular systems, which could be useful for  a
better understanding of their dynamics as well as for the design of
various control strategies using incoherent fields.
\cite{Dunbar,Chelk,Rabitz}.

\begin{acknowledgments}
We thank Anatole Kenfack for illuminating discussions during our joint time at the Max Planck Institute for the Physics of Complex Systems in Dresden.
\end{acknowledgments}


\begin{thebibliography}{99}

\bibitem{Zewail}  
 E. D. Potter {\em et al.}, Nature {\bf 355}, 66 (1992).

 \bibitem{Assion}  
 A. Assion {\em et al.}, Science {\bf 282}, 919 (1998).

\bibitem{RabitPhysTod} 
  Ian Walmsley and H. Rabitz, Phys. Today, Aug. {\bf 43} (2003).

\bibitem{AttoScience} 
 P. B. Corkum and F. Krausz, Nature Physics {\bf 3}, 381 (2007).

\bibitem{Johnsson}  
 P. Johnsson {\em et al.}, Phys. Rev. Lett {\bf 99}, 233001 (2007).

\bibitem{AttoSpect} 
 P. H. Bucksbaum, Science {\bf 317}, 766 (2007).

\bibitem{Rabitz-1}  
 A. Pechen and H. Rabitz   
 Phys. Rev. A {\bf 73}, 062102 (2006).

\bibitem{kenny} 
 A. Kenfack and J. M. Rost, J. Chem. Phys. {\bf 123}, 204322 (2005).

 \bibitem{kamal_prl}  
 Kamal P. Singh and Jan M. Rost
 Phys. Rev. Lett. {\bf 98}, 160201 (2007).

\bibitem{NoiseIonRyd} 
 R. Blumel {\em et al.}, Phys. Rev. Lett. {\bf 62}, 341 (1989);
 J. G. Leopold and D. Richards, J. Phys. B {\bf 24}, L243 (1991);
 L. Sirko {\em et al.}, Phys. Rev. Lett. {\bf 71}, 2895 (1993).

\bibitem{Rabitz-2}  
 F. Shuang and H. Rabitz 
 J. Chem. Phys. A {\bf 124}, 154105 (2006).

\bibitem{milner}
X. G. Xu, S. O. Konorov, J. W. Hepburn, and V. Milner, Nature
Physics {\bf 4}, 125 (2008).

\bibitem{Gamat} 
 L. Gammaitoni, P. H\"anggi, P. Jung, and F. Marchesoni,
 Rev. Mod. Phys. {\bf 70}, 223 (1998).

\bibitem{Buchl} 
 T. Wellens, V. Shatokhin, and A. Buchleitner,
 Rep. Prog. Phys. {\bf 67}, 45 (2004).

\bibitem{QSR} 
 R. L\"ofstedt and S. N. Coppersmith,
 Phys. Rev. Lett {\bf 72}, 1947 (1994).

\bibitem{chaoLgt} 
 D. Yelin, D. Meshulach, and Y. Silberberg, Opt. Lett. {\bf 22}, 1793 (1997).

 \bibitem{MIRshaping} 
 T. Witte, D. Zeidler, D. Proch, K. L. Kompa, and M. Motzkus
 Opt. Lett. {\bf 27}, 1793 (2002).

 \bibitem{Weiner} 
 A. M. Weiner, Rev. Scien. Instrument {\bf 71}, 1929 (2000).

\bibitem{HHG} 
 P. Agostini and L. F. DiMauro, Rep. Prog. Phys. {\bf 67}, 813 (2004).

\bibitem{Eberly}  
 M. Protopapas {\em et al.}, Rep. Prog. Phys. {\bf 60}, 389 (1997);
 J. Javanainen, J. H. Eberly and Q. Su, Phys. Rev. A {\bf 38}, 3430 (1988).

\bibitem{mpi}  
 G. Mainfray and C. Manus, Rep. Prog. Phys. {\bf 54}, 1333 (1991).

 \bibitem{kamal}  
 Kamal P. Singh, A. Kenfack and Jan M. Rost
 Phys. Rev. A {\bf 77}, 022707 (2008).

\bibitem{kamal_pra}  
 Kamal P. Singh and Jan M. Rost
 Phys. Rev. A {\bf 76}, 063403 (2007).

\bibitem{SOper}  
 J. A. Fleck, J. R. Moris, and M. D. Feit, Appl. Phys. {\bf 10}, 129 (1976).

\bibitem{QMbook}  
 C. Cohen-Tannoudji, B. Diu, and F. Lalo\"e, Quantum Mechanics, Vol. I (Springer-Verlag, Berlin, 1983).

\bibitem{Dunbar} 
 R. C. Dunbar and T. B. McMahon, Science {\bf 279}, 194 (1998).

\bibitem{Chelk} 
 S. Chelkowski {\em et al.}, Phys Rev. Lett. {\bf 65}, 2355 (1990);
 A. Assion {\em et al.}, Science {\bf 282}, 919 (1998).

\bibitem{Rabitz} 
  H. Rabitz {\em et al.}, Science {\bf 288}, 824 (2000).



\end{thebibliography}
\end{document}